\RequirePackage{lineno}
\documentclass[aps,prc,showpacs,twocolumn,superscriptaddress]{revtex4}
\usepackage{multirow}
\usepackage{graphicx}
\usepackage{amsmath}
\usepackage{lineno}
\usepackage{color}
\usepackage{amsmath}
\usepackage{amssymb}
\usepackage{MnSymbol}
\usepackage{subfigure}

\usepackage[normalem]{ulem} 

\def\version{version 3.0}

\newcommand{\pt}{$p_{T}$}

\begin{document}
\title{
\begin{flushright}
{
\small \sl \version \\
}
\end{flushright}

A study of charm quark correlations in ultra-relativistic $p$ + $p$ collisions with PYTHIA
}
\author{Shusu Shi}\affiliation{Key Laboratory of Quarks and Lepton Physics (MOE) and Institute of Particle Physics, Central China Normal University, Wuhan, 430079, China}
\author{Xin Dong}\affiliation{Lawrence Berkeley National Laboratory, Berkeley, California 94720, USA}
\author{Mustafa Mustafa}\affiliation{Lawrence Berkeley National Laboratory, Berkeley, California 94720, USA}

\date{\today}

\begin{abstract}

Abstract: Measurements of heavy flavor quark (charm and bottom) correlations in heavy ion
collisions are instrumental to understand the flavor dependence of energy
loss mechanisms in hot and dense QCD media.  Experimental measurements
 of these correlations in baseline $p$+$p$ collisions are crucial
to understand the contributions of perturbative and non-perturbative QCD
processes to the correlation functions and further help in interpreting
correlation measurements in heavy ion collisions. In this paper, we investigate
$D$-$\bar{D}$ meson correlations and $D$ with one particle from $D$ meson
decay daughter correlations using PYTHIA
Event Generator in $p$ + $p$ collisions at $\sqrt{s}$ = 200, 500 and
5500 GeV. Charm/bottom events are
found to contribute mainly to the away side/near side pattern of
$D$-electron correlations,
respectively. In the energy region of RHIC, $D$-$\bar{D}$ correlations
inherit initial $c$-$\bar{c}$ correlations and 
$B\rightarrow DX$ decay contribution is
insignificant. Furthermore, Bottom quark correlations are
suggested to be applicable at LHC energy, as the bottom contributions on
$D$ related correlations are relatively large. 
\end{abstract}
\maketitle
\clearpage
\section{Introduction}
\label{sect_intro}

Experimental observables indicate that strongly coupled
Quark-Gluon Plasma (sQGP) is formed in the high energy
heavy ion collisions~\cite{rhicwp1, rhicwp2, rhicwp3, rhicwp4}. Heavy quarks ($c$ and $b$) serve
  as good probes to study the properties of sQGP, as charm and bottom can only
be pair-produced in early stage hard scatterings due to the large
mass ($>$ 1 GeV/$c^{2}$)~\cite{hf_lin}.  Experimental
  measurements have shown that heavy quarks suppression at relatively high-\pt\
is as large as that of light quarks~\cite{nonpe1, nonpe2}, which was inconsistent with earlier theoretical expectation of the
flavor dependence of parton energy loss. Thus it is crucial to understand
the charm-medium interaction mechanism.  Recently, the azimuthal
correlations of heavy quarks are found to have the potential for
distinguishing different energy loss mechanisms inside the hot
medium~\cite{cc_theory1, cc_theory2}.  The theoretical prediction indicates
that pure radiative energy loss
does not change the initial angular correlation function in a significant
way, whereas pure collisional energy loss is
more efficient at diluting initial back-to-back charm pair correlation,
this could even lead to a peak in the near side at
low-\pt\ ~\cite{xianglei2}.  To approach the azimuthal correlations of heavy
quarks in experiment, measurement of $D$-$\bar{D}$ correlations is
the ideal candidate~\cite{xianglei1, xianglei2}, as
$D$-$\bar{D}$ correlations inherit most of charm pair
correlations. However, due to the relatively small charm
cross section, small hadronic decay branching ratio which is used to
reconstruct $D$ meson and limited signal to background ratio, the direct
measurement of $D$-$\bar{D}$ correlations would be very challenging. One has
to consider use $D$-$X$ or $X$-$D$ correlation measurement instead of direct
$D$-$\bar{D}$, where $X$ is a decay daughter
of $D$ meson. 

Considering only leading order perturbative Quantum Chromodynamics (pQCD)
processes, the initial charm pairs exhibit exactly 
back-to-back correlation.  High order pQCD processes broaden the initial
charm correlations.  Charm quarks finally
fragment into charmed hadrons, the non-perturbative fragmentation processes
further broaden the correlations.  In the case of $D$-$X$ or $X$-$D$
correlation, decay kinematics and random combinatorial pairs would dilute or
even destroy the correlation signal originating from initial charm pair
correlations.   It is crucial to understand these effects on the correlation
function firstly in $p$ + $p$ collisions which serve as the baseline for heavy
ion collisions.  Furthermore, an experimental measurement
of $D$-$\bar{D}$ and $D$-$X$ ($X$-$D$) correlations in $p$ + $p$ collisions
would constrain the contributions from high order processes of pQCD
calculations and non-perturbative fragmentation processes.  In this paper, we
utilize PYTHIA Event Generator (EG) to study $D$-$\bar{D}$
and $D$-$X$ ($X$-$D$) correlations in $p$ + $p$ collisions at Relativistic
Heavy Ion Collider (RHIC) and Large Hadron Collider (LHC) energies, $\sqrt{s} $ = 200, 500 and 5500
GeV. 

The rest of the article is organized in the following way: Section II summarizes the details of tuning PYTHIA parameters. 
In Sec. III, we present the correlations results of $D$-$\bar{D}$ and $D$-$X$ ($X$-$D$). The comparsion is made beween
RHIC and LHC energies. Last a summary is presented in Sec. IV.

\section{PYTHIA tune}
\label{tune}

\begin{figure*}[ht]
\vskip 0cm
\includegraphics[width=1.0\textwidth]{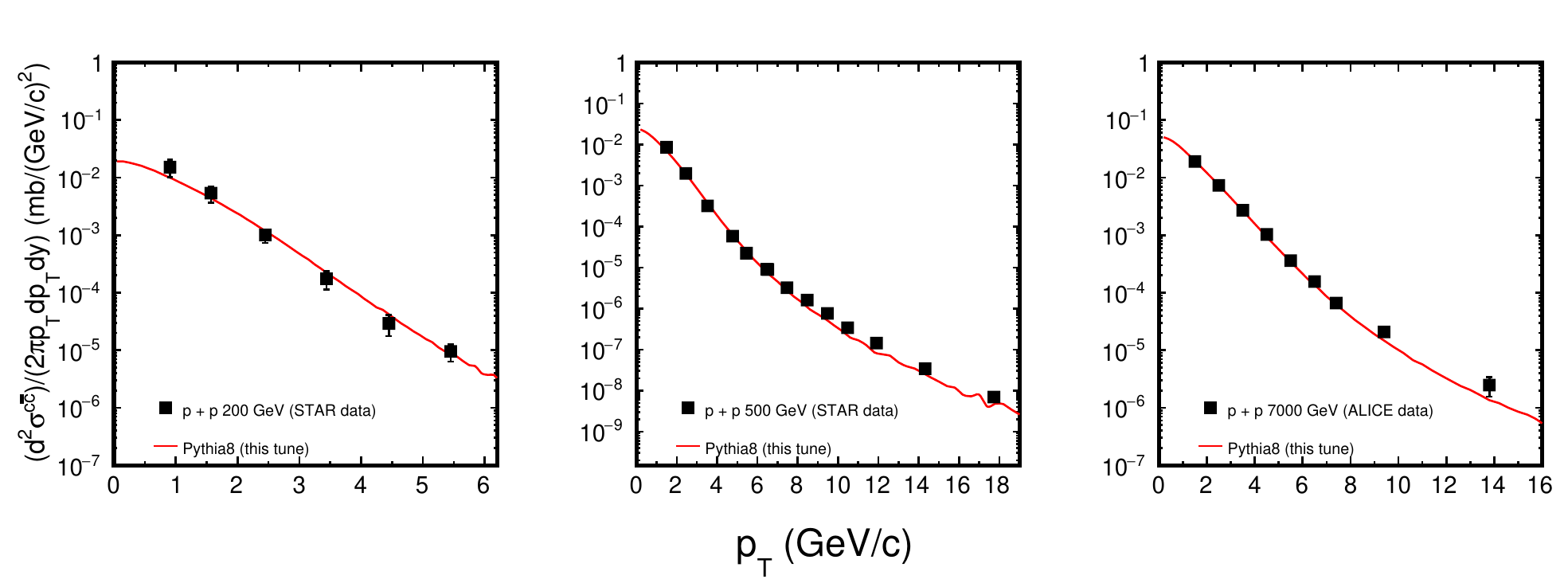}
\caption{ (Color online) $c$$\bar{c}$ production cross section as a function of transverse momentum in $p$ + $p$ collisions
  at $\sqrt{s}$ = 200, 500 and 7000 GeV from STAR and ALICE experimental data (solid squares) and tuned PYTHIA (red lines).
} \label{fig:pythia_tune}
\end{figure*}

PYTHIA8 (version8.168) has been used in this study. In the following, we will
use PYTHIA representing PYTHIA8. We tuned PYTHIA parameters to match
the $c\bar{c}$ production cross-section as inferred from measurement of
$D^{0}$ and $D^{*}$ production in $p$ + $p$ collisions at $\sqrt{s}$ =
200 and 500 GeV measured in the STAR experiment at
RHIC~\cite{starcc_200, starcc_500} and
7000 GeV measured in the ALICE experiment at LHC~\cite{alicecc_7000}.
The choice of modifying the strong
interaction coupling constant ($\alpha_{s}$) value of final parton shower
(TimeShower:alphaSvalue) and minimum invariant transverse momentum ($p_{T}$)
threshold for hard QCD process (PhaseSpace:pTHatMin) gives the best
$\chi^{2}/{\rm ndf}$ to the data.
Figure 1 shows the $c\bar{c}$ production cross section as a function of $p_T$
from STAR measurement and PYTHIA.  In the PYTHIA calculation, all ground state
charm hadrons ($D^{0}$, $\bar{D^{0}}$, $D^{+}$, $\bar{D^{-}}$, $D^{+}_{s}$,
$\bar{D^{+}_{s}}$, $\Lambda^{+}_{c}$ and $\bar{\Lambda^{-}_{c}}$) were added
together in the rapidity window $|y| < 1$ to obtain charm cross-section.  The
parameters of `TimeShower:alphaSvalue' and `PhaseSpace:pTHatMin' were set to
0.18 and 1.3 GeV/$c$ for $\sqrt{s}$ = 200 GeV collisions.   These two parameters were set to 0.15 and 1.5 GeV/$c$ for $\sqrt{s}$ =
500 GeV collisions; 0.15 and 2.8 GeV/$c$ for 7000 GeV.  
The $\chi^{2}/{\rm ndf}$ values are 1.31 for 200 GeV data, 2.13 for 500 GeV data and 1.95 for 7000 GeV data,
respectively. As lack of $p$ + $p$ experimental data in top energy of heavy ion collisions at LHC (5500 GeV), 
the same setup of parameters as 7000 GeV was applied in PYTHIA simulation.

\section{Correlations}

\begin{figure*}[ht]
\vskip 0cm
\includegraphics[width=0.8\textwidth]{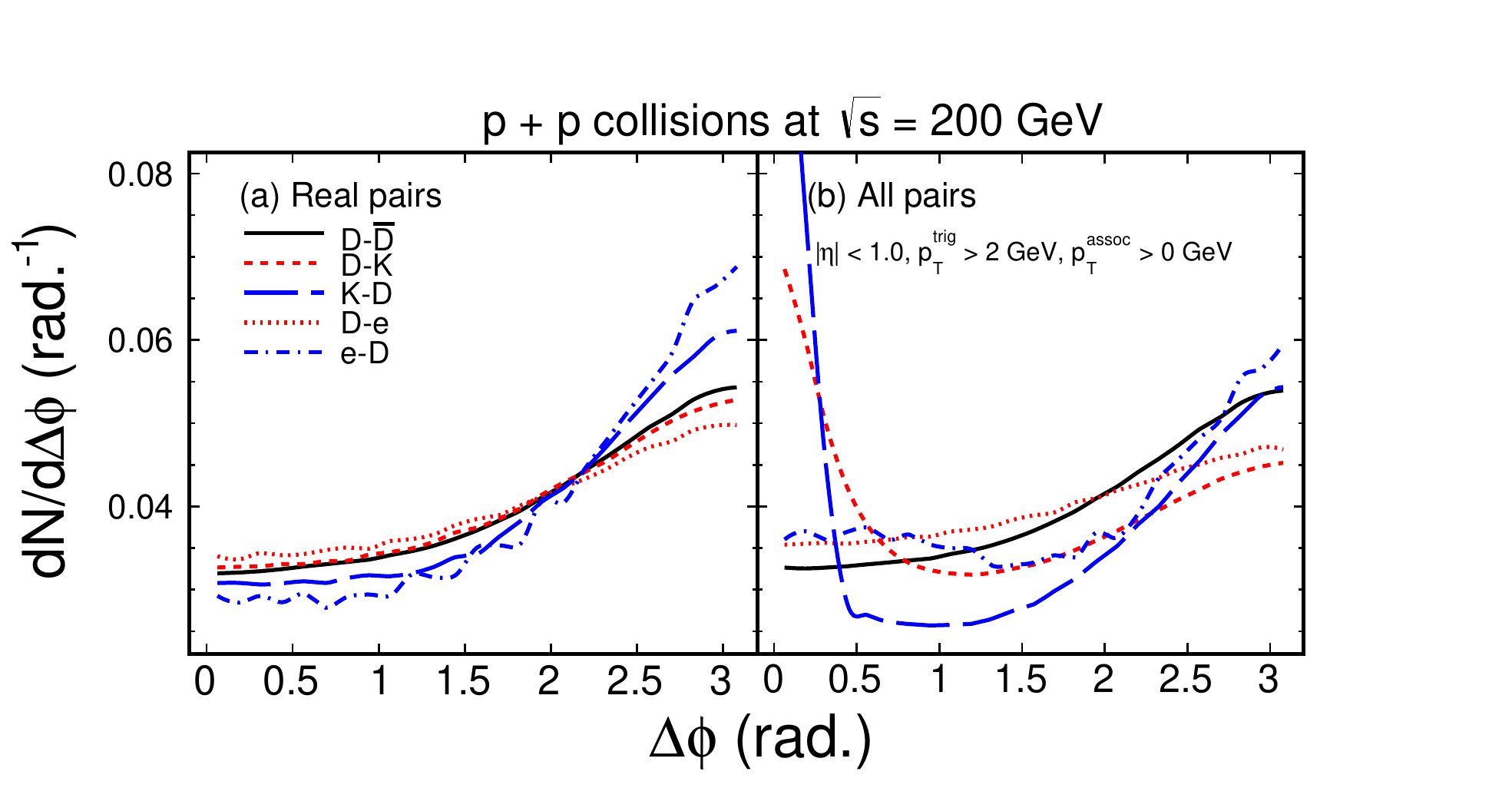}
\caption{ (Color online) The $D$-$\bar{D}$, $D$-$K$, $K$-$D$, $D$-$e$ and $e$-$D$ correlations as a function of
  relative azimuth angle $\Delta \phi$ in $p$ + $p$ collisions at $\sqrt{s}$ = 200 GeV in PYTHIA for (a) real pairs (pairs are from $c$-$\bar{c}$ pairs), 
  b) all pairs (all possible pairs in an event). The phase space cut is pseudorapidity ($|\eta| < 1$). An
  additional $p_T$ cut is applied to trigger particles. The integrals of correlation functions are normalized to one.
} \label{fig:correlation_200}
\end{figure*}

\begin{figure*}[ht]
\centering
\subfigure[]{%
\includegraphics[scale=0.4]{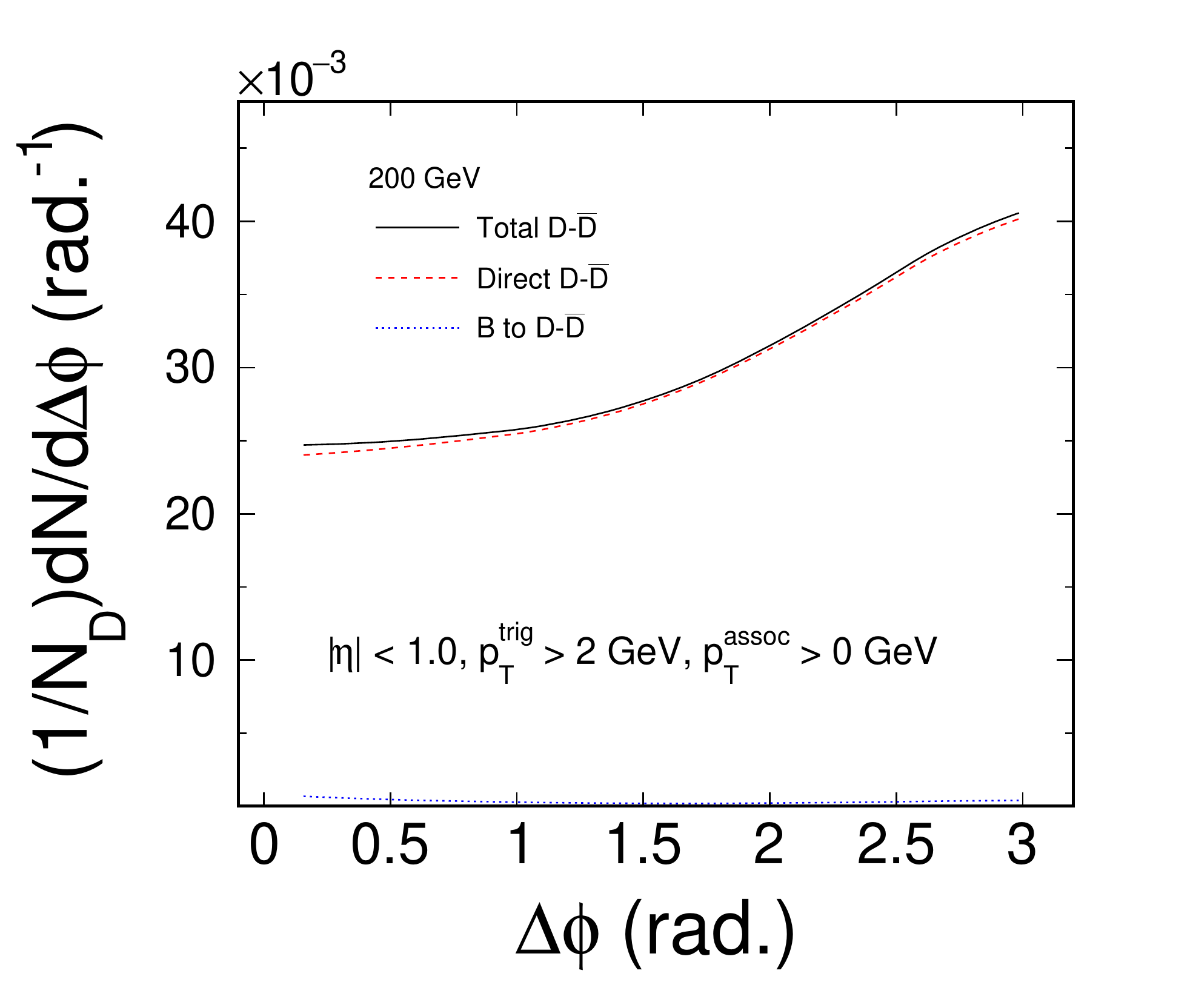}
\label{fig:subfigure1}}
\quad
\subfigure[]{%
\includegraphics[scale=0.4]{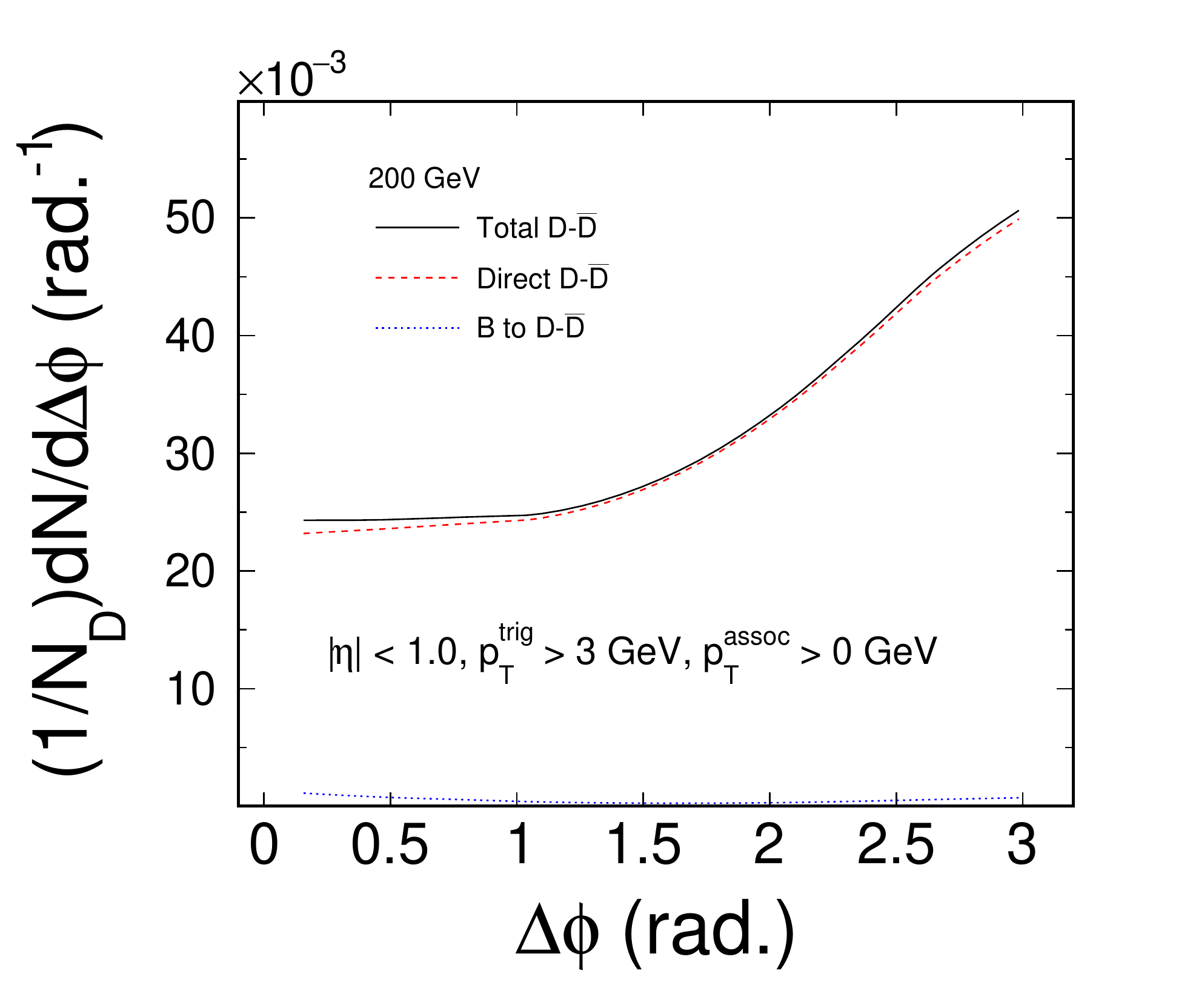}
\label{fig:subfigure2}}

\subfigure[]{%
\includegraphics[scale=0.4]{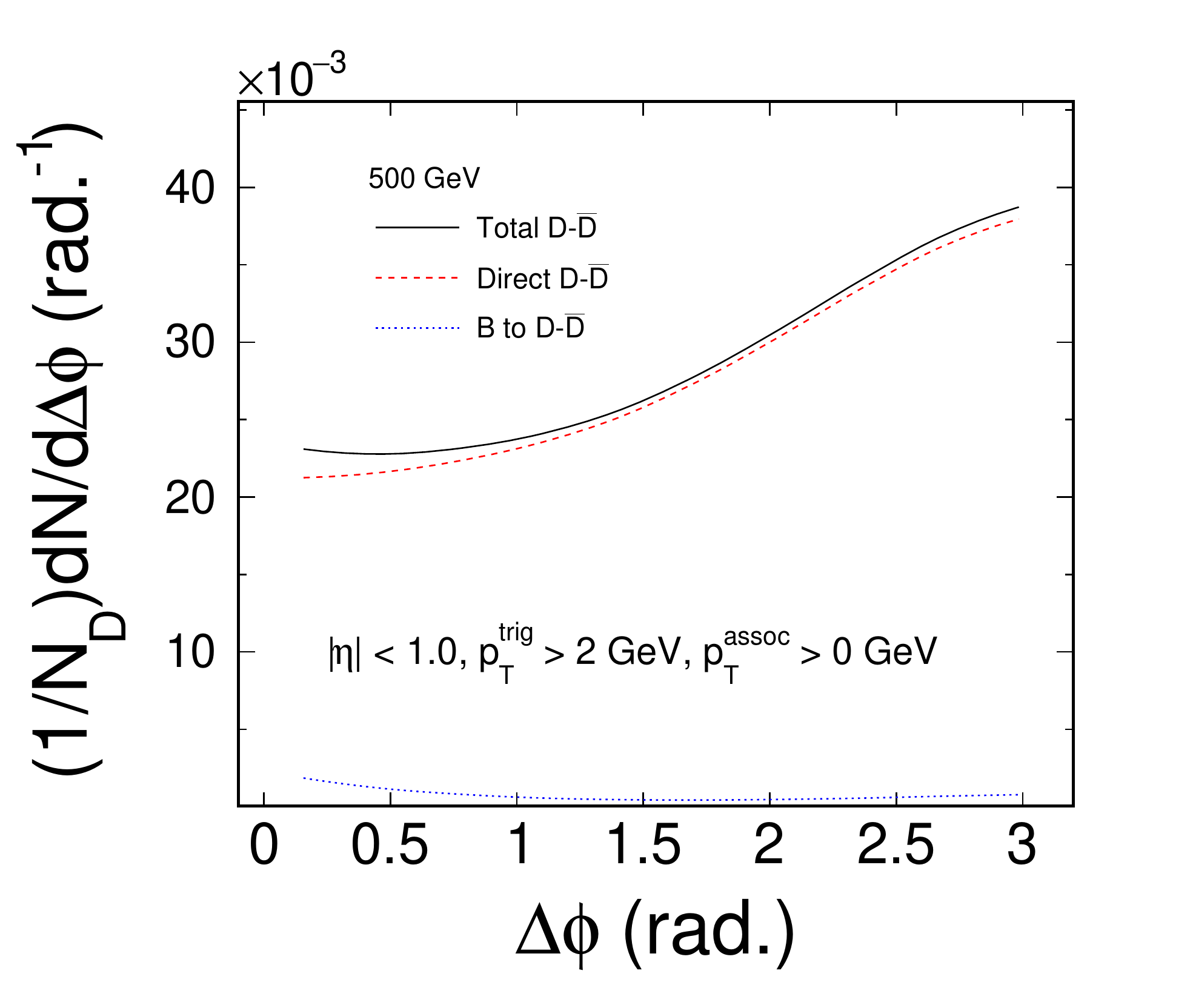}
\label{fig:subfigure3}}
\quad
\subfigure[]{%
\includegraphics[scale=0.4]{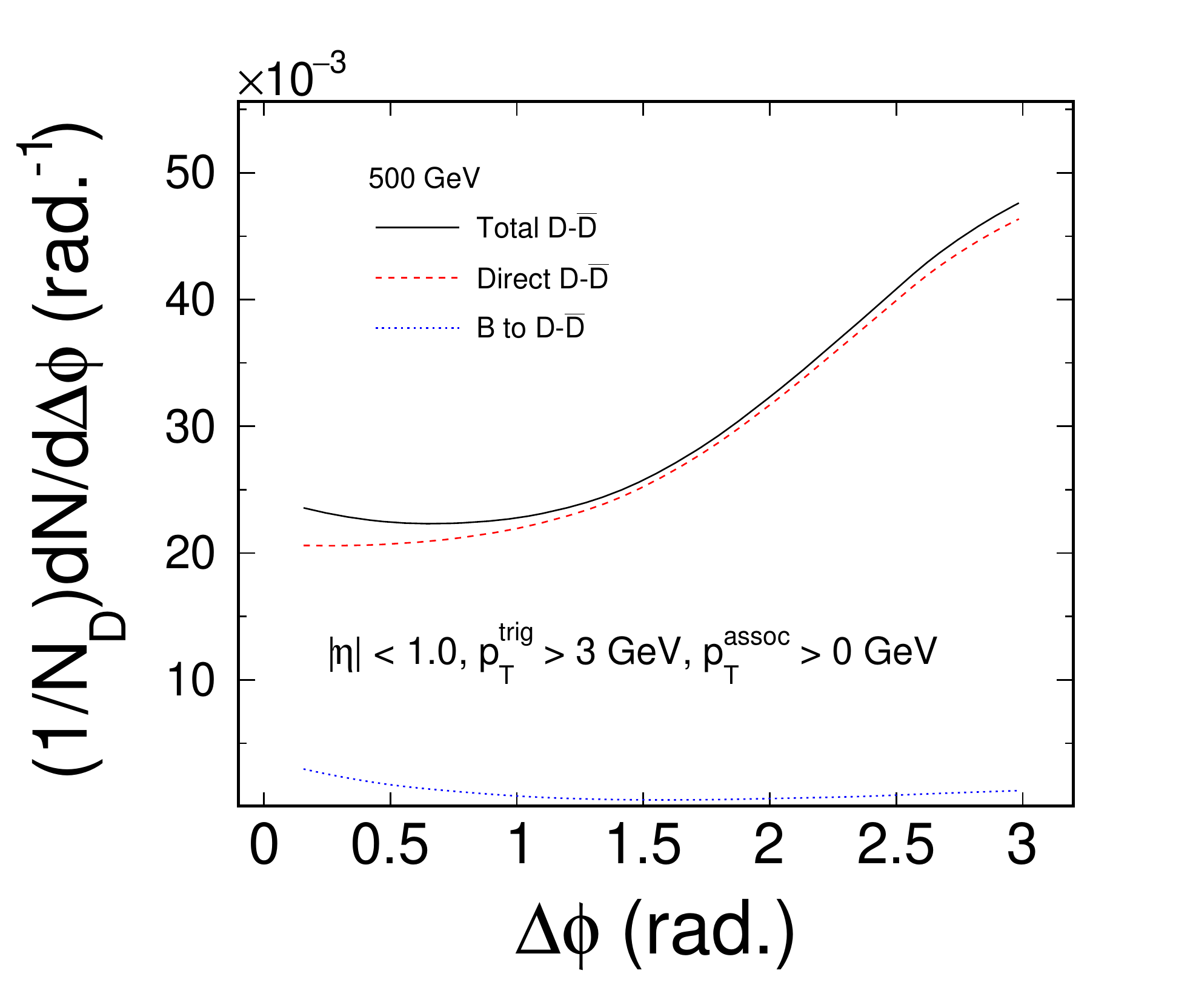}
\label{fig:subfigure4}}
\caption{(Color online) The $D$-$\bar{D}$ correlations from different sources as a function of relative azimuth angle $\Delta \phi$ in $p$ + $p$ collisions at $\sqrt{s}$ = 200 GeV 
(panel (a) and (b)) and 500 GeV (panel (c) and (d)). 
The minimum trigger $p_T$ cuts are set to 2 GeV/$c$ (panel (a) and (c)) and 3 GeV/$c$ (panel (b) and (d)). The correlation functions are normalized by
the number of $D$ mesons.}
\label{fig:DD_correlation}
\end{figure*}
\begin{figure*}[ht]
\centering
\subfigure[]{%
\includegraphics[scale=0.7]{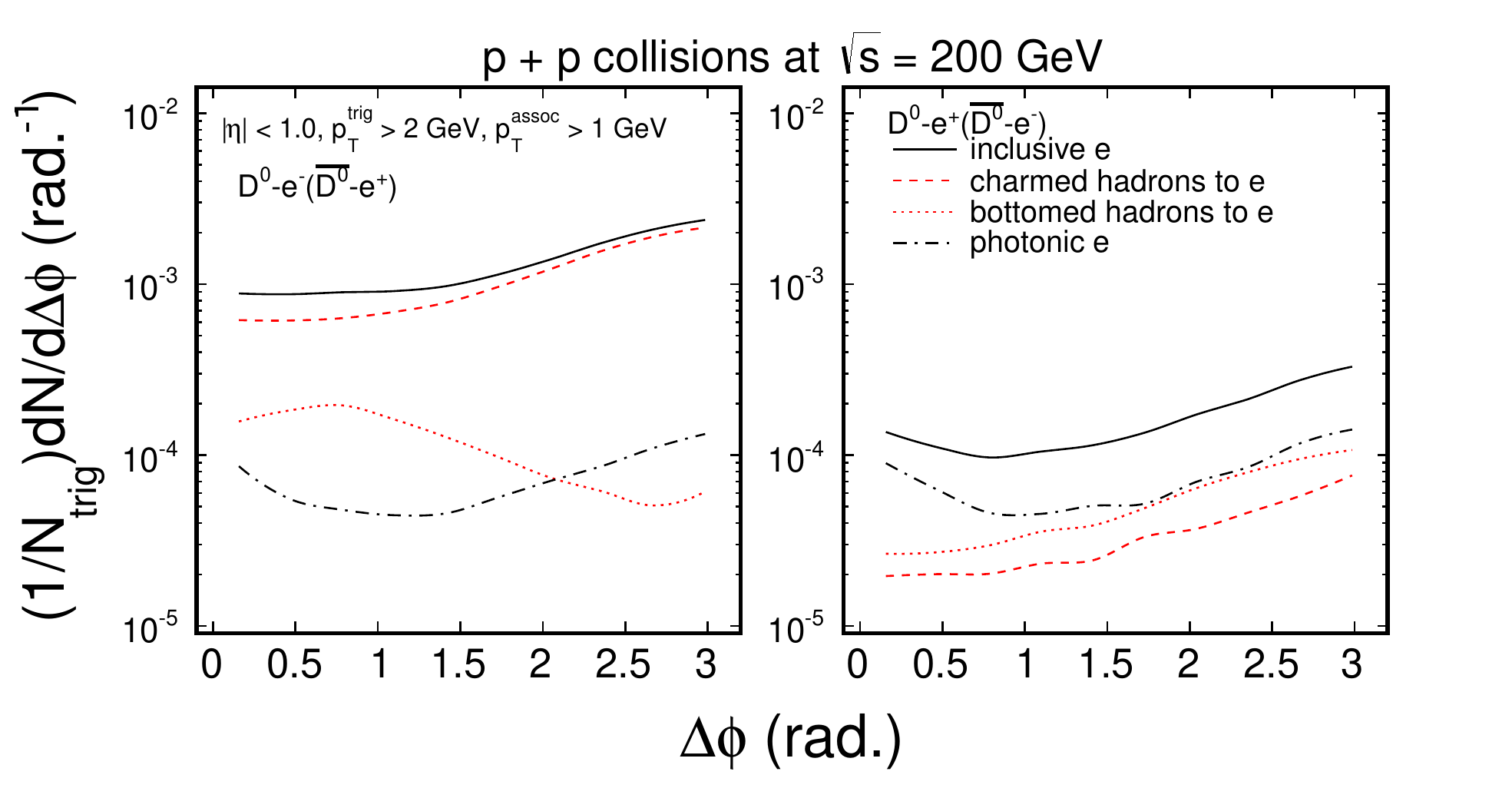}
\label{fig:subfigure1}}
\subfigure[]{%
\includegraphics[scale=0.7]{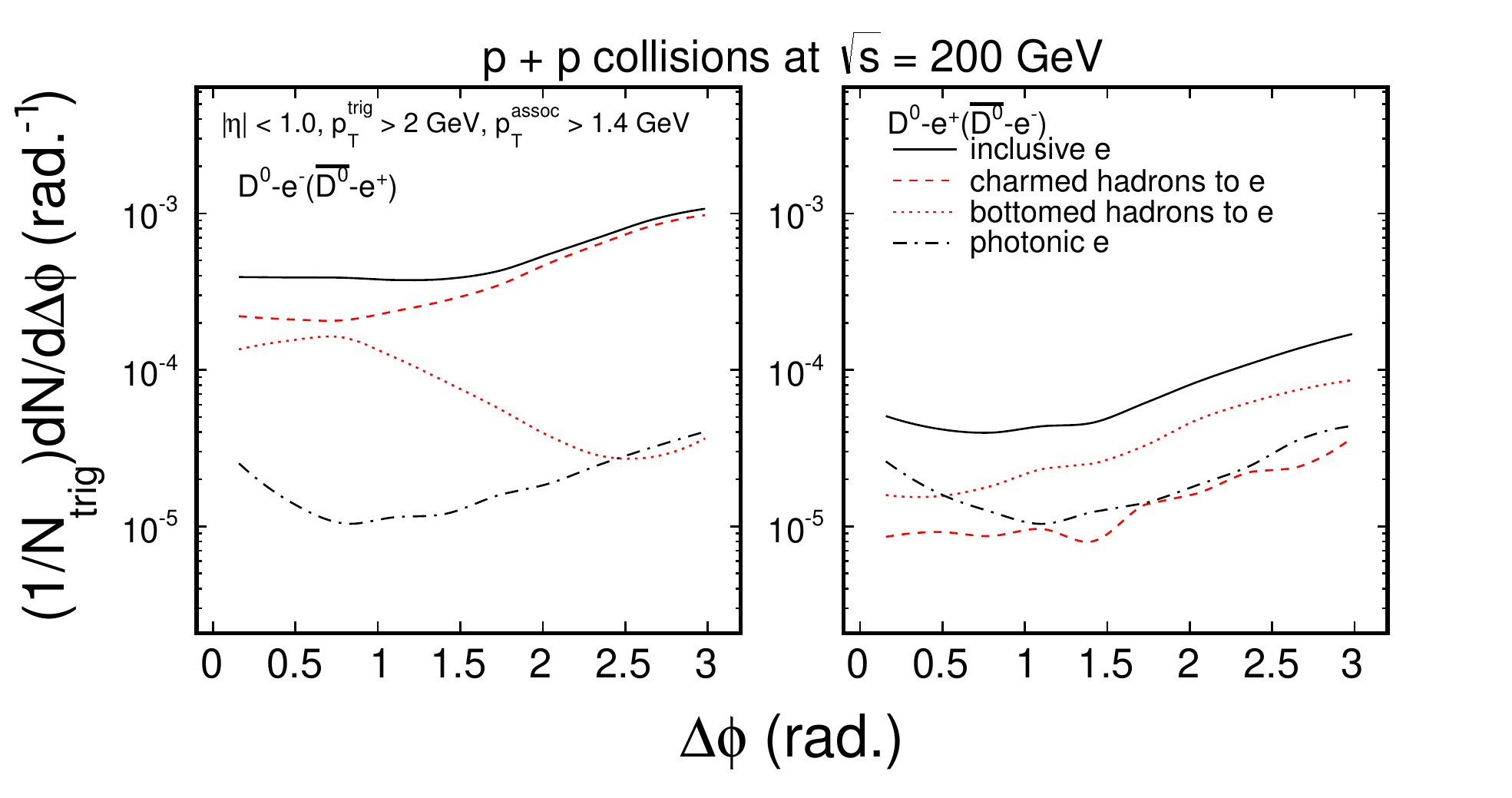}
\label{fig:subfigure2}}
\caption{(Color online) The $D^{0}$-$e^{-}$($\bar{D^{0}}$-$e^{+}$)  and $D^{0}$-$e^{+}$($\bar{D^{0}}$-$e^{-}$) correlations
as a function of relative azimuth angle $\Delta \phi$ in $p$ + $p$ collisions at $\sqrt{s}$ = 200 GeV. Different lines represent
differnt sources of electron sample.  The minimum trigger $p_T$ cut is 2 GeV/$c$ and the minimum associate $p_T$ cut is 1 GeV/$c$ for panel (a) and 1.4 GeV/$c$ for panel (b), respectively.}
\label{fig:De_correlation}
\end{figure*}

\begin{figure*}[ht]
\vskip 0cm
\includegraphics[width=0.8\textwidth]{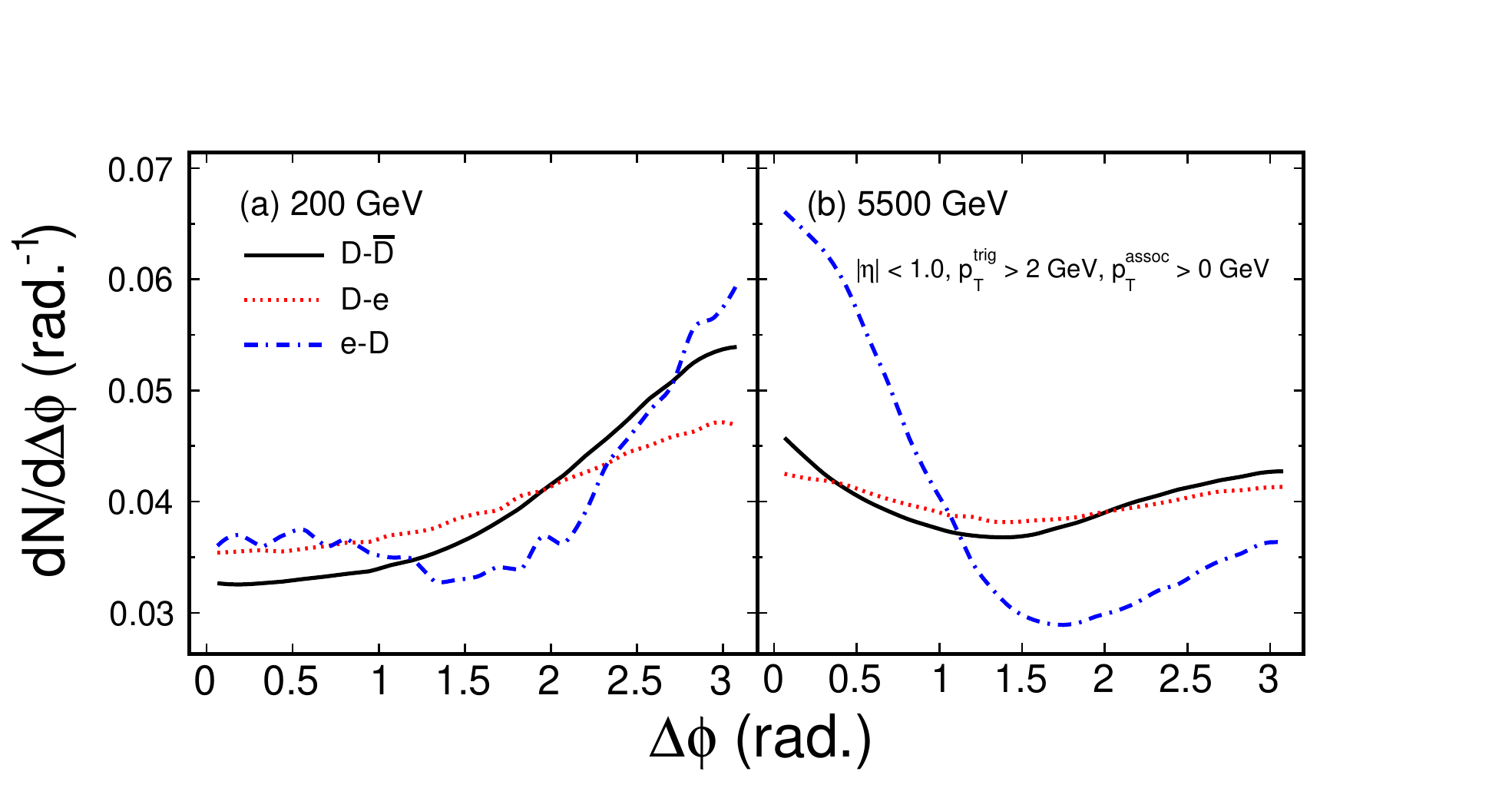}
\caption{ (Color online) The $D$-$\bar{D}$, $D$-$e$ and $e$-$D$ correlations as a function of
relative azimuth angle $\Delta \phi$ in $p$ + $p$ collisions calculated by PYTHIA at (a) $\sqrt{s}$ = 200 GeV and (b) 5500 GeV. The integral of correlation function are normalized to one.
} \label{fig:rhic_lhc}
\end{figure*}

\begin{figure*}[ht]
\vskip 0cm
\includegraphics[width=0.6\textwidth]{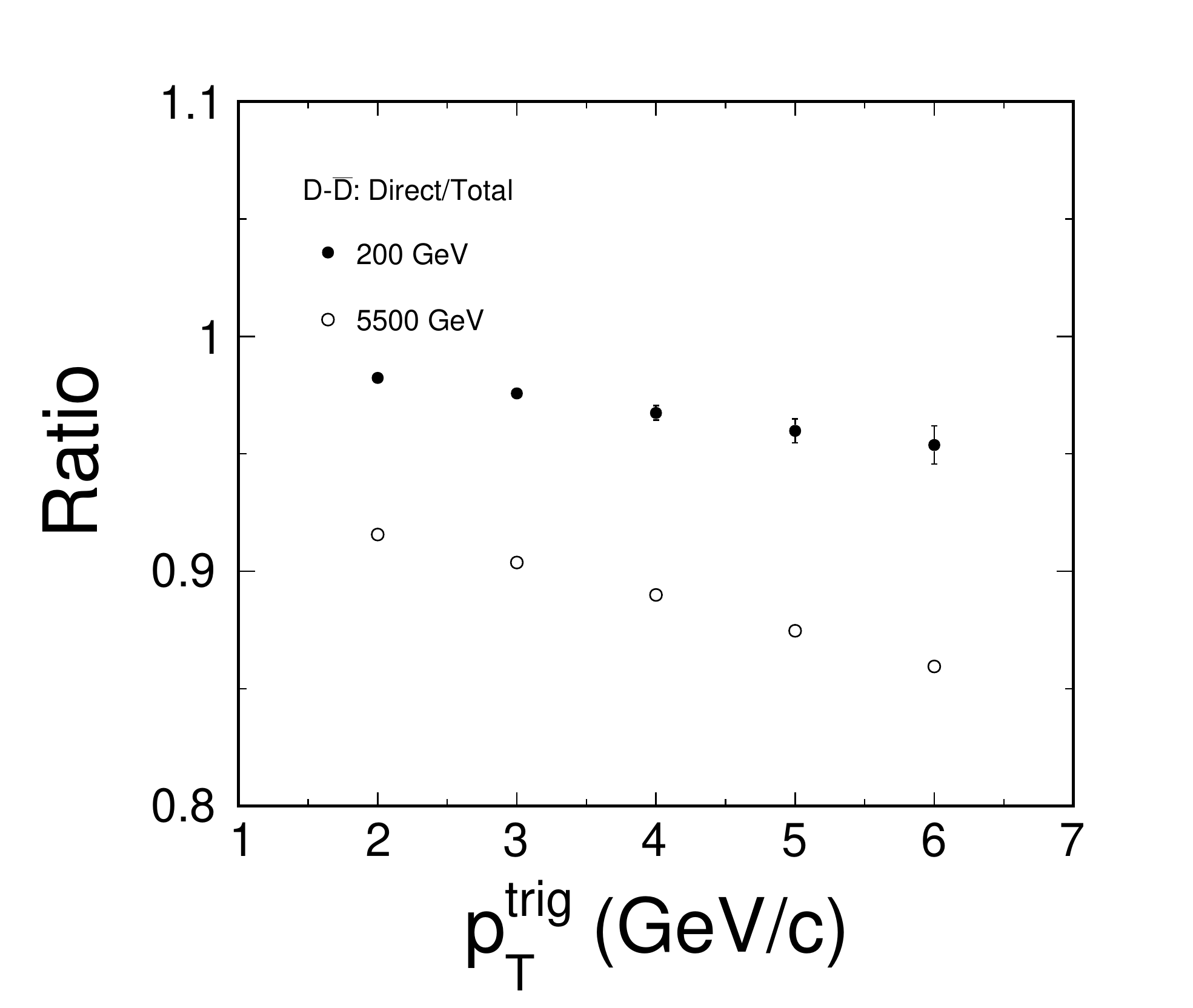}
\caption{The ratio of direct $D$-$\bar{D}$ pairs to total $D$-$\bar{D}$ pairs as a function of trigger transverse momentum in $p$ + $p$ collisions at $\sqrt{s}$ = 200 and 5500 GeV.} \label{fig:DD_correlation}
\end{figure*}

$D$-$\bar{D}$ correlations are the ideal
experimental way to study the azimuthal correlations of initial $c$-$\bar{c}$
pairs, as the charm quark mostly fragment to $D$ mesons. In this paper,
we use $D$ to represent the combination of $D^{0}$, $D^{+}$, $D^{+}_{s}$ and
$\bar{D}$ to represent $\bar{D^{0}}$, $\bar{D^{-}}$, $\bar{D^{+}_{s}}$.  In
experiment, $D$ mesons can be reconstructed by the hadronic decay channels.
Figure 2 shows the $D$-$\bar{D}$ correlations in $p$ + $p$ collisions at
$\sqrt{s}$ = 200 GeV for (a) real pairs and (b) all pairs. Where real pairs
mean $D$-$\bar{D}$ from $c$-$\bar{c}$ pairs, all pairs
mean all possible $D$-$\bar{D}$ pairs in an event.  A phase space cut of
pseudorapidity ($|\eta| < 1$) was applied to match the typical STAR
experimental acceptance. The integral of all the correlation functions were
normalized to one, as we are interested in the correlation patterns here. The
$D$-$\bar{D}$ azimuthal correlation function exhibits a clear away side
correlation inherited from the initial $c$-$\bar{c}$ pair
production. In case we only consider the leading order (LO) pQCD contribution,
the  $c$-$\bar{c}$ would show a delta function like away side correlations. The
next-to-leading (NLO) order pQCD contributions including radiative corrections,
flavor excitation and gluon splitting broadens the correlation function.  There
are no exact NLO pQCD calculations in PYTHIA, but the
initial and final parton shower procedure emulates these effects. Thus
some aspects of the multiple parton emission phenomenon could be well
reproduced~\cite{pythia_highorder1, pythia_highorder2}. The non-perturbative
fragmentation process further broadens the correlation function. In 
other words, the correlation function of $D$-$\bar{D}$ inherits the initial
$c$-$\bar{c}$ back-to-back correlation broadened by higher order pQCD and fragmentation
processes. The results of all pairs in panel (b) are consistent with real pairs
in panel (a), because most events only contain one $D$-$\bar{D}$ pair with
 phase space cuts (($|\eta| < 1$)) in $p$ + $p$ collision at RHIC energies.  The experimental
measurement of  $D$-$\bar{D}$ correlations in $p$ + $p$ collisions is crucial
to constrain the higher order pQCD and fragmentation calculations. Further in
the heavy ion collision system, Au + Au for example, the initial away side
correlation is expected to by the charm-medium
interactions. Different energy loss mechanisms, collision energy loss or
radiative energy loss, show dramatically different
modification~\cite{cc_theory1, cc_theory2}. The experimental measurement of
$D$-$\bar{D}$ correlation in heavy ion collision system will help us to
disentangle the question on charm quark energy loss mechanism.  However, it is a quite
challenging analysis, the statistics are limited by the small $c$-$\bar{c}$
production cross section, the usable hadronic decay branch ratio (e. g. $D^{0}
\rightarrow K^{-} + \pi^{+}$) and the limited signal over background ratio of
reconstructed $D$ mesons~\cite{starcc_200}.

\subsection{$D$-$X$ ($X$-$D$) correlations}

Instead of direct $D$-$\bar{D}$ analysis, $D$-$X$ or $X$-$D$ correlations are
other options to approach  charm quarks correlations, where $D$-$X$
means correlations of $D$($\bar{D}$) meson and a decay daughter of
$\bar{D}$($D$) meson (electrons, kaons, pions and so on). (First letter
indicates the trigger particle.) The typical choices would be charged kaons
and electrons. Charged pions are not good candidates, because
of the copious production and resonance decay contributions to pions.
Electrons could be divided into photonic and non-photonic electrons. Photonic
electrons include those from $\gamma$ conversion and Dalitz decay, while
non-photonic electrons include those from charmed and bottomed
hadrons semileptonic decays~\cite{nonpe1, nonpe2}. Experimentally,
non-photonic electrons can be statistically subtracted
from inclusive electrons~\cite{nonpe1, nonpe2}. Thus in the PYTHIA simulation results in
Fig. 2, we use  non-photonic electrons for $D$-$e$ and $e$-$D$
correlations. Also shown in Fig. 2 are $D$-$K$ and $K$-$D$ correlations. Panel
(a) shows the correlation function inheriting from initial $c$-$\bar{c}$ pairs.
The away side correlations can be seen for all combinations. For $D$-$\bar{D}$,
$D$-$K$ and $D$-$e$, the different widths of away side peak are due to the
decay smearing of $D \rightarrow K$ or $D\rightarrow e$.  The different widths
of away side peak of $D$-$\bar{D}$, $K$-$D$ and $e$-$D$ are because minimum
trigger $p_{T}$ cut on $D$ decayed particles ($>$ 2 GeV/$c$)  actually requires
higher $p_{T}$ $D$ mother trigger particles. Panel (b) shows the correlation
function for all possible pairs in an event. The large near side peak for
$D$-$K$ and $K$-$D$ correlation are mainly due to the jet correlation, as most
kaons are not from $D$ decays. It suggests $D$ - inclusive-hadron correlations
are not good candidates to study the charm correlations. $e$-$D$ correlation
also shows a near side correlation pattern, 
which is due to the semileptonic decay of $B$ contribution. The STAR
measurement indicates that ratio of non-photonic electrons
from $B$ to those from $D$ is $~$0.2 at $p_T$ = 2 GeV/$c$~\cite{star_BD}. The
random pairs which do not originate from the same
$c$-$\bar{c}$ pair cause additional combinatorial background.  It
is partly the reason why the away side correlation width of $D$-$e$ is wider
than that of $D$-$\bar{D}$. We also found tighter or looser cut on minimum transverse momentum of 
associate particle does not change the
correlation patterns dramatically.

\subsection{Components of $D$-$\bar{D}$ correlations}
\label{DD}

The $D$($\bar{D}$) mesons consist of directly produced $D$($\bar{D}$) and
$D$($\bar{D}$) from $B$($\bar{B}$) meson decays.  To study the $B$ contribution to
 the $D$-$\bar{D}$ correlation function, we separate
the $D$-$\bar{D}$ pair combination into four cases: 1) direct $D$ and
$\bar{D}$, 2) $D$ from $B$ decays and $\bar{D}$ from $B$ decays,
3) $D$ from $B$ decays and direct $\bar{D}$, 4) direct $D$, and $\bar{D}$ from
$B$ decays. The combination of case 2) + 3) +4) is generally called
$D$-$\bar{D}$ pairs from $B$ decays. Figure 3 shows the decomposed
$D$-$\bar{D}$ correlations in $p$ + $p$ collisions at $\sqrt{s}$ = 200 GeV ((a)
and (b))  and 500 GeV ((c) and (d)) for two sets of
 trigger particle minimum $p_T$ cut. It can be observed
that the $B$ decay contribution to $D$-$\bar{D}$ depends on trigger particle
$p_T$ cut and collision energy.  The higher trigger particle $p_T$ cut or
collisions energy, the larger contribution from $B$ decay. In comparison to
direct $D$-$\bar{D}$ pairs, $D$-$\bar{D}$ pairs from $B$ decays are
insignificant.  In $p$ + $p$ collisions at $\sqrt{s}$ = 500
GeV, with $p_{T}^{\rm trig} > $ 3 GeV/$c$ cut, the $D$-$\bar{D}$ pairs from $B$
decays are $\sim$ 4\% of the total $D$-$\bar{D}$ pairs. As shown in the panel
(d) of fig. 3, the $B$ contributions cause the small near side correlation
pattern of the total $D$-$\bar{D}$ correlation. In $p$ + $p$  collisions at
$\sqrt{s}$ = 200 GeV, even with $p_{T}^{\rm trig} > $ 3 GeV/$c$ cut, the
contribution from $B$ decays is still not sizable. It is
also found that tighter or looser cut on minimum transverse momentum of 
associate particle does not
change the conclusion. It indicates the $D$-$\bar{D}$ correlations mostly
inherit from $c$-$\bar{c}$ pair production at RHIC energies. The effect of $B$
decays on $D$-$\bar{D}$ correlations is insignificant.

\subsection{$D$-$e$ correlations}

$D$-$\rm electron$ ($D$-$e$) correlations contain contribution from different
sources, thus it is more complicated than $D$-$\bar{D}$
correlation.  The inclusive electrons consist of photonic electrons and
non-photonic electrons. The photonic electrons are from $\gamma$ conversion (in
experiment) and Dalitz decay (e. g. $\pi^{0}$, $\eta$), where the non-photonic
electrons are from semileptonic decay of charmed and bottomed
hadrons. The non-photonic electrons can be statistically separated from
the inclusive electron samples in experiment~\cite{nonpe1, nonpe2} which makes
the $D$ and non-photonic electron correlation measurement possible. Figure 4
shows the $D^{0}$-$e$ correlations from different electron sources
in PYTHIA.   Following the typical
experimental way, we chose $D^{0}$ from golden hadronic
decay channel $D^{0} \rightarrow K^{+} + \pi^{-}$. Based on the sources of
electrons, we plot the the
correlation of $D^{0}$: 1) to electrons from
charmed hadrons, 2) to electrons from
bottomed hadrons, 3) to photonic electrons. Note 
that charmed hadrons contain the direct charm and
decay contributions from bottomed hadrons, to
be consistent with the experimental capability.  The inclusive
electrons are simply the sum of all three cases mentioned above.  As the low
$p_T$ electrons are mainly from photonic sample, a minimum $p_T$ cut is usually
applied on electrons.  Panel (a) and (b) show the minimum $p_T$ cut of
electrons = 1 and 1.4 GeV/$c$ respectively.  The initial $c$-$\bar{c}$
correlations contribute to $D^{0}$-$e^{-}$ and $\bar{D^{0}}$-$e^{+}$
correlations which are shown in the left panels, where $D^{0}$-$e^{+}$ and
$\bar{D^{0}}$-$e^{-}$ can be interpreted as other contributions which are shown
in the right panels. It is clear that the near side correlation pattern of
$D^{0}$-$e^{-}$ is mainly from bottom events and the away side correlation
pattern of $D^{0}$-$e^{-}$ is mainly from charm events. As discussed in
reference~\cite{correlation_method}, the azimuthal correlation of $D^{0}$ and
non-photonic electrons allows the separation of charm and bottom production on
a statistical basis. The $D^{0}$ - photonic $e^{-}$ and $D^{0}$ - photonic
$e^{+}$ correlation pattern are symmetric, because the Daliz decay or the
$\gamma$ conversion contributes equally to $e^{+}$ and $e^{-}$. As
the minimum $p_T$ cut on electrons increases, the
bottom contributions become larger. This can be explained by the increase of $B
\rightarrow e$ to $D \rightarrow e$ ratio as a function of
$p_T$~\cite{{star_BD}}. The results of $p$ + $p$ at $\sqrt{s}$ = 500 GeV is
similar to the 200 GeV results.  To investigate the $c$-$\bar{c}$ correlations
by measuring $D$-$e$ correlation, one has to stick to away side region which is
dominant by charm contributions. The $\Delta\phi$ region of 2 - $\pi$ would be
fine inferring from PYTHIA simulation.

\subsection{RHIC versus LHC energy}
From RHIC to LHC, as the collision energy increases, higher order pQCD and bottom
contributions to the correlation functions become more
and more significant. In Fig. 5, we compare the $D$-$\bar{D}$, $D$-$e$
(non-photonic electrons) and $e$-$D$ correlations in $p$ + $p$ collisions at
$\sqrt{s}$ = 200 and 5500 GeV.  We chose these two collisions energies, as they
are the top heavy ion collision energies at RHIC and LHC, respectively.  
One can observe the near side correlations show almost the same
magnitude as the away side correlations for $D$-$\bar{D}$ correlation
function at the LHC energy. Furthermore, the higher order pQCD processes
smear the away side correlations significantly: in the $D$-$e$ case, the
correlation function is almost flat in $p$ + $p$ collisions at $\sqrt{s}$ =
5500 GeV.  The large near side peak of $e$-$D$ correlations are from $B$ to $e$
contributions. 
Similar as Sec. III B, we separate the $D$-$\bar{D}$
correlations into direct $D$ and $B$ to $D$ correlations.  It is found the near
side correlations are from $B$ to $D$ contributions.  Quantitatively, in Fig. 6, we show the ratio of direct 
$D$-$\bar{D}$ pairs to total $D$-$\bar{D}$ for 200 and 5500 GeV.
It is observed that the $D$-$\bar{D}$ pairs from $B$ contributions are relatively larger in LHC energy than RHIC energy.
Up to trigger $p_T$ cut at 6 GeV/$c$, the $D$-$\bar{D}$ pairs from $B$ contributions are less than 5\% in 200 GeV
collisions where the $B$ contribution is $\sim$15\% in 5500 GeV collisions.
The PYTHIA results suggest that charm
correlations might not be the best choice at the LHC energy, as the baseline measurements in $p$ + $p$ are significantly
affected by $B$ contributions and higher order pQCD. On the other hand, bottom correlations, especially $B$-$\bar{B}$
correlations are clean probes for both $p$ + $p$ and heavy ion
collisions and possible to be done at LHC energies, due to
the larger production cross section.

\section{Summary}
\label{sect_summary}

With PYTHIA (version 8.168) framework, we studied $D$-$\bar{D}$ and $D$-$X$
($X$-$D$) correlations in $p$ + $p$ collisions at $\sqrt{s}$ = 200 and 500 GeV.
$D$-hadron correlation is found to be effected significantly by the jet
correlations. $D$-$\bar{D}$ correlation is the ideal choice
to approach initial $c$-$\bar{c}$ pair correlations. The bottom decay
contributions to $D$-$\bar{D}$ pairs are less than 4\% at $p$ + $p$ collisions
in $\sqrt{s}$ = 500 GeV with trigger $p_T$ cut at 3 GeV/$c$, thus it is
not significant at RHIC energies. $D$ - non-photonic electron correlation contains
both contributions from charm and bottom events. The charm/bottom events
dominate the near/away side correlation pattern respectively.  $D$ -
non-photonic electron correlations can be used to determine the branching
ratios for the charm and bottom decays to electrons.  One can also study
$c$-$\bar{c}$ correlations from the away side pattern of $D$-$e$ correlation
with a proper $\Delta \phi$ cut.  The charm quark correlations are found to be
less attractive at LHC energies, due to the relatively larger bottom and higher order
pQCD contributions. Alternatively, the bottom quark correlations
would be applicable at LHC.  Experiment measurements on heavy quark ($c$ and
$b$) correlations are crucial to understand the charm correlations in $p$ + $p$
collision system, thus offering precious data to constrain the pQCD
calculations. Furthermore, in heavy ion collisions, these measurements are
important to investigate the charm-medium interactions and disentangle the
contributions of different charm energy loss mechanisms in the hot and dense
medium. 

\section{Acknowledgments}
This work was supported in part by the National Natural Science Foundation of China under grant No. 11475070
and National Basic Research Program of China (973 program) under grand No. 2015CB8569.

%

%
\end{document}